\documentclass[preprint, showpacs, aps, draft]{revtex4}
\usepackage{bm}
\begin{document}
\title{Bosonic field equations from an exact uncertainty principle
\footnote{Journal
     of Physics A: Mathematical and General \copyright 
(2003) IOP
     Publishing Ltd.}}
\author{Michael J. W. Hall$^\ast$}
\author{Kailash Kumar$^\ast$}
\author{Marcel Reginatto$^\dagger$}
\affiliation{{}$^\ast$Theoretical Physics, IAS, \\ Australian National
University,\\
Canberra ACT 0200, Australia}
\affiliation{{}$^\dagger$Physikalisch-Technische Bundesanstalt, \\
Bundesallee 100\\
38116 Braunschweig, Germany}
\begin{abstract}
A Hamiltonian formalism is used to describe ensembles of fields in terms
of two canonically conjugate functionals (one being the field
probability density).  The postulate that a classical ensemble is
subject to nonclassical fluctuations of the field momentum density, of a
strength determined solely by the field uncertainty, is shown to lead to a
unique modification of the ensemble Hamiltonian. The modified equations
of motion are equivalent to the quantum equations for a bosonic field,
and thus this exact uncertainty principle provides a new approach to
deriving and interpreting the properties of quantum ensembles.  
The examples of electromagnetic and gravitational fields are discussed.  
In the latter case the exact uncertainty approach specifies a unique
operator ordering for the Wheeler-DeWitt and Ashtekar-Wheeler-DeWitt equations. 
\end{abstract}
\maketitle

\renewcommand{\thesection}{\arabic{section}}

\section{Introduction}

The Heisenberg uncertainty relation
\begin{equation} \label{hur}
\Delta x \,\Delta p \geq \hbar/2
\end{equation}
for the rms position and momentum uncertainties of a quantum particle is
a well known feature of quantum mechanics. It has recently 
been shown, however, that this relation is a consequence of a more
fundamental connection between the statistics of complementary quantum
observables.

In particular, the distinguishing ``nonclassical'' 
property of complementary observables is
that they cannot be simultaneously measured to an arbitrary accuracy.
It is therefore natural to consider the decomposition of one such
observable, the momentum say, into the sum of a ``classical'' and a
``nonclassical'' component: 
\[ \hat{p} = \hat{p}^{cl} + \hat{p}^{nc}~, \]
where the classical component, $\hat{p}^{cl}$, is defined as that observable
{\it closest} to $\hat{p}$ (in a statistical sense) which {\it is} 
simultaneously
measurable with the complementary observable $\hat{x}$ \cite{eur}.  

It turns out that
such decompositions do indeed, in a number of ways, 
neatly separate classical and
nonclassical properties of quantum observables.  
For example, for one-dimensional particles the {\it
nonclassical} component of the momentum satisfies the uncertainty
relation 
\begin{equation} \label{eur}
\delta x \,\Delta p^{nc} = \hbar/2
\end{equation}
for {\it all} pure states, where $\delta x$ denotes 
a measure of position uncertainty from
classical statistics called the Fisher length \cite{eur}.
This {\it exact} uncertainty relation is far stronger than
(and implies) the corresponding Heisenberg uncertainty relation in Eq.
(\ref{hur}). 

The surprising fact that quantum particles satisfy exact uncertainty
relations has recently provided the basis for {\it deriving} much of the quantum
formalism, from an exact uncertainty principle.  In particular, the
assumption that a classical ensemble is subjected to nonclassical momentum
fluctuations, of a strength inversely proportional to uncertainty in
position, has been shown to lead directly from the classical equations
of motion to the Schr\"{o}dinger equation \cite{hallreg}. A brief
overview of exact uncertainty properties of quantum particles is
provided by 
the conference paper in Ref.~\cite{bamberg}. 

The aim of this paper is to show that the exact uncertainty principle
may be successfully generalised to derive the equations of motion for
bosonic fields with Hamiltonians quadratic in the field
momenta (eg, scalar, electromagnetic and gravitational fields).  
This ``exact uncertainty'' approach is extremely minimalist in nature:
unlike canonical quantisation, it does
not use nor make any assumptions about
the existence of operators, Hilbert spaces, complex
amplitudes, inner products, 
linearity, superposition, or the like. The sole ``nonclassical''
element needed is the addition of fluctuations to the
momentum density of members of a classical ensemble of fields, 
with the fluctuation statistics assumed to be determined by the ensemble
statistics. The exact uncertainty approach is thus 
conceptually very simple, being based
on the core notion of statistical uncertainty 
(intrinsic to any interpretation of quantum theory). 
As a bonus, the
exact uncertainty approach further
implies a unique operator ordering for the Schr\"{o}dinger 
equation associated with the quantum ensemble - something
which the canonical quantisation procedure is unable to do.  

It is remarkable
that the basic underlying concept - the addition of ``nonclassical'' momentum
fluctuations to a classical ensemble - carries through from quantum
particles to quantum fields, without creating
conceptual difficulties (although significant technical generalisations
are needed). This logical consistency and range of applicability 
is a further strength of the exact uncertainty 
approach.

In the next section the equations of motion for a classical ensemble of
fields are expressed in Hamiltonian form, via two canonically conjugate
functionals (the probability density and the Hamilton-Jacobi functional).
In Sec.~3 it is shown that the classical ensemble Hamiltonian is modified
by the addition of nonclassical momentum density fluctuations, in a
manner uniquely specified by the exact uncertainty principle, 
leading to modified equations of motion equivalent to the quantum field
equations. 

In Secs.~4 and 5 the examples of the electromagnetic and
gravitational fields are discussed.  In the former case it is shown
that the exact uncertainty approach is equivalent to adding nonclassical
fluctuations to the electric field ${\bf E}$, of a strength determined
by the inherent uncertainty of the 
vector potential ${\bf A}$.  In the latter case it is shown that
the exact uncertainty approach leads to a unique operator ordering for
the Wheeler-DeWitt and Ashtekar-Wheeler-DeWitt 
equations.   This ordering is, moreover, 
consistent with
Vilenkin's ``tunneling" boundary condition for inflationary cosmology
\cite {vilenkin}.   
Results are discussed in Sec.~6, and necessary elements from
classical field theory and functional analysis, and the proof of the
main Theorem of Sec.~3, are given in
the Appendices.
 
\section{Classical ensembles}

We first consider a real multicomponent classical field $f\equiv(f^a)$ 
with conjugate
momentum density $g\equiv(g^a)$, described by some Hamiltonian functional
$H[f,g,t]$. For example, $f$ may denote the electromagnetic field
$A\equiv(A^\mu)$, or some collection of interacting fields labelled by the
index $a$. 
Spatial coordinates will be denoted by $x$ (irrespective of dimension), and
the values of field components $f^a$ and $g^b$ at position $x$ will be 
denoted by $f^a_x$ and $g^b_x$ respectively.

There are three canonical approaches to describing the evolution of an
{\it individual} field, 
based on the (related) Lagrangian, Hamiltonian, and Hamilton-Jacobi
formalisms respectively. We choose the latter here, as it provides a
straightforward mechanism for adding
 momentum fluctuations to an {\it ensemble} of
fields (Eq.~(\ref{noise}) below).  The Hamilton-Jacobi formalism also
leads to an elegant ``ensemble Hamiltonian'' 
representation for the dynamics of an ensemble of 
fields, in terms of two canonically conjugate functionals 
(Eq.~(\ref{mconj}) below).

First, the equation of motion for an individual classical field 
is given by the Hamilton-Jacobi equation
\begin{equation} \label{mhj}
\frac{\partial S}{\partial t} + H[f,\delta S/\delta f, t] = 0 ,
\end{equation}
where $S[f]$ denotes the Hamilton-Jacobi functional, and $\delta/\delta f$
denotes the functional derivative with respect to $f$ (see Appendices A
and B). The momentum density associated with field $f$
is given by $g=\delta S/\delta f$, and hence $S$ will also be referred
to as the {\it momentum potential}. 

Second, the description of an {\it ensemble} of
such fields further requires a probability density functional, $P[f]$.
The equation of motion for $P[f]$ corresponds to the conservation of
probability, i.e., to the continuity equation
\begin{equation} \label{mcont}
\frac{\partial P}{\partial t} + \sum_a\int dx\, \frac{\delta}{\delta
f^a_x}\left(P\left.\frac{\delta H}{\delta g^a_x}\right|_{g=\delta
S/\delta f}\right) = 0 ,
\end{equation}
as is reviewed in Appendix B. 

Eqs.~(\ref{mhj}) and (\ref{mcont}) describe the motion of the ensemble
completely, in terms of the two functionals $P$ and $S$. 
These equations of motion may be written in the ``Hamiltonian'' form 
\begin{equation} \label{mconj}
\frac{\partial P}{\partial t} = \frac{\Delta\tilde{H}}{\Delta S}, ~~~~~
\frac{\partial S}{\partial t} = - \frac{\Delta\tilde{H}}{\Delta P} , 
\end{equation}
where $\tilde{H}$ denotes the functional integral
\begin{equation} \label{mham}
\tilde{H}[P, S, t]:= \langle H\rangle = \int D\!f\, PH[f, \delta
S/\delta f, t] ,
\end{equation}
and where variational derivatives such as $\Delta\tilde{H}/\Delta S$ are
discussed in Appendix A. The equivalence of Eqs. (\ref{mhj})
and (\ref{mcont}) to Eqs. (\ref{mconj}) follows directly from Eq.
(\ref{mform}).
 
The functional integral $\tilde{H}$ in Eq.~(\ref{mham}) will 
therefore be referred to as the {\it ensemble Hamiltonian}, and, 
in analogy to Eqs. (\ref{conj}) 
of Appendix B, $P$ {\it and} $S$ {\it may be regarded as
canonically conjugate functionals}.
Note from Eq. (\ref{mham}) that $\tilde{H}$ typically corresponds to the
mean energy of the ensemble.  It may be shown that Eqs.~(\ref{mconj}) 
follow from the action principle
$\Delta\tilde{A}=0$, with action $\tilde{A} = \int dt [-\tilde{H} +
\int D\!f\,S (\partial P/\partial t)]$.

In what follows, we specialise to ensembles for which the
associated 
Hamiltonian functional is  {\it quadratic} in the momentum field
density, i.e., of the form
\begin{equation} \label{hquad}
H[f, g, t] = \sum_{a,b}\int dx\, K^{ab}_{x}[f] g^a_xg^b_x + V[f] .
\end{equation}
Here $K^{ab}_x[f]=K^{ba}_x[f]$ is a kinetic factor coupling components
of the momentum density, and $V[f]$ is some potential energy functional. 
The corresponding
ensemble Hamiltonian is given by Eq. (\ref{mham}).  Note that cross terms
of the form $g^a_xg^b_{x'}$ with $x\neq x'$
are not permitted in local field theories, and
hence are not considered here. 

\section{Momentum fluctuations $\Rightarrow$ quantum ensembles}

The ensemble Hamiltonian, $\tilde{H}[P,S,t]$ in Eq. (\ref{mham}), 
is our classical
starting point for describing an ensemble of fields.  This starting
point must be modified in some way if one is to obtain new equations of
motion, to be identified as describing a {\it quantum} 
ensemble of fields.  
Our approach to modifying the ensemble Hamiltonian is based on a single  
ingredient: the addition of nonclassical fluctuations to the momentum
density, with the magnitude of the fluctuations determined by the
uncertainty in the field. This ``exact uncertainty'' approach leads to
equations of motion equivalent to those of a bosonic field,  
with the interpretational advantage of an intuitive  
statistical picture for quantum field ensembles, and the
technical advantage of a unique operator ordering for the
associated Schr\"{o}dinger equation.

Suppose then that $\delta S/\delta f$ is in fact an {\it average}
momentum density associated with field $f$, in the sense that the true momentum
density is given by 
\begin{equation} \label{noise}
g = \delta S/\delta f + N,
\end{equation}
where $N$ is a fluctuation field that vanishes on the average for any
given field $f$. Thus the physical meaning of $S$ changes to being an
{\it average} momentum potential. 
No specific underlying model for $N$ is assumed or necessary: in the approach
to be followed, one may in fact
interpret the ``source'' of the fluctuations as the field
uncertainty itself. Thus the main effect of the fluctuation field 
is to remove any deterministic connection between $f$
and $g$. 

Since the momentum fluctuations may conceivably depend on the field $f$,
the average over such fluctuations for a given quantity $A[f,N]$ will be
denoted by $\overline{A}[f]$, and the average over fluctuations {\it
and} the field by $\langle A\rangle$.  Thus $\overline{N}\equiv 0$ 
by assumption, and in general $\langle A\rangle = \int
D\!f\,P[f]\, \overline{A}[f]$.  Assuming a quadratic dependence on
momentum density as per Eq. (\ref{hquad}), it follows 
that when the fluctuations are
significant the classical 
ensemble Hamiltonian $\tilde{H} = \langle H\rangle$ in Eq.~(\ref{mham}) should
be replaced by
\begin{eqnarray}
\tilde{H}' & = & \langle~ H[f, \delta S/\delta f + N, t]~\rangle\nonumber\\\
& = & \sum_{a,b}\int D\!f\,\int dx\, P K^{ab}_x \overline{( \delta
S/\delta f^a_x +N^a_x) (\delta S/\delta f^b_x +N^b_x)} + \langle
V\rangle\nonumber\\ \label{hprime}
& = & \tilde{H} + \sum_{a,b}\int D\!f\, \int dx\, P K^{ab}_x \,\overline{
N^a_xN^b_x} .  
\end{eqnarray}
Thus the momentum fluctuations lead to an additional nonclassical term in the 
ensemble Hamiltonian, specified by the {\it covariance matrix} ${\rm
Cov}_x(N)$ of the fluctuations at position $x$, where
\begin{equation} \label{covn}
\left[{\rm Cov}_x(N)\right]^{ab} := \overline{N^a_xN^b_x} .  
\end{equation}

The covariance matrix is uniquely determined, up to a
multiplicative constant, by the following four assumptions:

{\it (1) Causality:}  $\tilde{H}'$ is an ensemble Hamiltonian for the
canonically conjugate functionals $P$ and $S$, which yields causal equations
of motion. Thus no higher than first-order functional derivatives can appear in
the additional term in Eq.~({\ref{hprime}), implying that 
\[
{\rm Cov}_x(N) = \alpha(P, \delta P/\delta f_x, S, \delta S/\delta
f_x, f_x, t) 
\]
for some symmetric matrix function $\alpha$.
Note that in principle one could also allow the covariance matrix to depend on
auxilary fields and functionals; however, the fourth assumption below
immediately removes such a possibility.

{\it (2) Independence:}  If the ensemble comprises two independent
non-interacting subensembles 1 and 2, with a factorisable probability
density functional $P[f^{(1)},f^{(2)}] = P_1[f^{(1)}] P_2[f^{(2)}]$, 
then any dependence of the corresponding subensemble fluctuations 
$N^{(1)}$ and $N^{(2)}$ on $P$ only enters via
the corresponding probability densities $P_1$ and $P_2$ respectively. Thus
\[
\left.{\rm Cov}_x(N^{(1)})\right|_{P_1P_2} = \left.{\rm
Cov}_x(N^{(1)})\right|_{P_1}, ~~~~\left.{\rm
Cov}_x(N^{(2)})\right|_{P_1P_2} =\left.{\rm Cov}_x(N^{(2)})\right|_{P_2}
\]
for such an ensemble.  
Note that this assumption implies that the ensemble Hamiltonian 
$\tilde{H}'$ in Eq.~(\ref{hprime}) is {\it additive} 
for independent non-interacting ensembles (as is the
corresponding action $\tilde{A}'$). 

{\it (3) Invariance:} The covariance matrix transforms correctly under
linear canonical transformations of the field components.  Thus, noting 
that $f\rightarrow\Lambda^{-1}f$, $g\rightarrow\Lambda^Tg$ is a
canonical transformation for any invertible matrix $\Lambda$ with
transpose $\Lambda^T$, 
which preserves the quadratic form of $H$
in Eq. (\ref{hquad}) and leaves the momentum potential $S$ invariant (since
$\delta/\delta f\rightarrow \Lambda^T\delta/\delta f$), one has from 
Eq. (\ref{noise}) that
$N\rightarrow\Lambda^TN$, and hence that
\[ {\rm Cov}_x(N) \rightarrow \Lambda^T {\rm Cov}_x(N)\Lambda ~~~{\rm for}
~~~f\rightarrow\Lambda^{-1}f .\]
Note that for single-component fields this reduces to a scaling relation
for the variance of the fluctuations at each point $x$.
  
{\it (4) Exact uncertainty:} The uncertainty of the momentum density
fluctuations at any given position and time, as characterised by the
covariance matrix of the fluctuations, 
is specified by the field uncertainty at that position and
time.  Thus, since the field uncertainty is completely determined by the
probability density functional $P$, it follows that ${\rm Cov}_x(N)$
cannot depend on $S$, nor explicitly on $t$. 

It is seen that the first three assumptions (causality, independence and
invariance) are natural on physical grounds, and hence relatively
unconstraining.  In contrast, the fourth assumption is ``special'': 
it postulates an exact connection
between the nonclassical momentum uncertainty and the field uncertainty.   
Remarkably, these assumptions lead
directly to equations of motion of a bosonic quantum field, as shown by
the following Theorem and Corollary.

{\bf Theorem:}  {\it The above assumptions of causality, independence,
invariance, and 
exact uncertainty imply that} 
\begin{equation} \label{theorem}
\overline{N^a_xN^b_x} = C (\delta P/\delta f^a_x) (\delta P/\delta
f^b_x)/P^2 ,
\end{equation}
{\it where $C$ is a positive universal constant.}  

The theorem thus yields a unique form for the additional term in Eq.
(\ref{hprime}), up to a multiplicative constant $C$.  The classical
equations of motion for the ensemble are recovered in the limit of small
fluctuations, i.e., in the limit
$C\rightarrow0$.  Note that one cannot make the identification 
$N^a_x\sim(\delta P/\delta f^a_x)/P$ from Eq. (\ref{theorem}),
as this is inconsistent with the fundamental property
$\overline{N^a_x}=0$.
The proof of the theorem is given in Appendix C, and is 
substantially different from (and stronger than) proofs of an analogous
theorem
for quantum particles \cite{hallreg, bamberg} (the latter proofs rely heavily
on a ``scalar'' assumption that does not carry over in a natural
manner to fields). 

The main result of this section is the following
corollary (proved in Appendix C):

{\bf Corollary:} {\it The equations of motion corresponding to the
ensemble Hamiltonian $\tilde{H}'$ can be expressed as the single complex
equation}
\begin{equation} \label{corollary}
i\hbar\frac{\partial\Psi}{\partial t} =
H[f,-i\hbar\delta/\delta f, t] \Psi = -\hbar^2 \left(\sum_{a,b}\int dx\,
\frac{\delta}{\delta f^a_x} K^{ab}_x[f] \frac{\delta}{\delta f^b_x}
\right) \Psi +
V[f]\Psi ,
\end{equation}
{\it where one defines}
\begin{equation} \label{hdef}
\hbar := 2\sqrt{C},~~~~~~\Psi := \sqrt{P} e^{iS/\hbar} .
\end{equation}

Eq. (\ref{corollary}) may be recognised as the Schr\"{o}dinger equation
for a quantum bosonic field \cite{brown, schweber}, and hence the goal of
deriving this equation, via an exact uncertainty principle for 
nonclassical
momentum fluctuations acting on a classical ensemble, has been achieved.  
Note that the
exact uncertainty approach specifies a {\it unique} operator ordering,
$(\delta/\delta f^a_x)K^{ab}_x(\delta/\delta f^b_x)$,
for the functional derivative operators in Eq. (\ref{corollary}).  Thus
there is no ambiguity in the ordering for cases where $K^{ab}_x$ depends on
the field $f$, in contrast to traditional
approaches (eg, the Wheeler-DeWitt equation, discussed in Sec.~5
below).  The above results generalise straightforwardly to complex
classical fields.  

The ensemble of fields corresponding to ensemble Hamiltonian
$\tilde{H}'$ will be called the {\it quantum ensemble}
corresponding to $\tilde{H}$. Note from Eqs.~(\ref{theorem}) and
(\ref{hdef}) that the role of Planck's constant is
to fix the relative scale of the nonclassical fluctuations.
It is remarkable that the four assumptions
of causality, independence, invariance and exact uncertainty lead to a
{\it
linear operator} equation.

Finally, it may be remarked that the equations of motion of a classical
ensemble may be subject to some imposed constraint(s) on $P$ and $S$. 
For example, each
member of an ensemble of electromagnetic fields may have the Lorentz gauge 
imposed (see Sec.~4
below). As a guiding principle, we will require that the corresponding
{\it quantum} ensemble is subject to the same constraint(s) on $P$ and $S$.
This will ensure a meaningful classical-quantum
correspondence for the results of field measurements.  However,
consistency of the quantum equations of motion with a given set of
constraints is not guaranteed by the above Theorem and Corollary, and so
must be checked independently for each case.

\section{Example: Electromagnetic field}

\subsection{Lorentz gauge}

The electromagnetic field is described, up to gauge invariance, 
by a 4-component field $A\equiv (A^\mu$).
In the Lorentz gauge all physical fields satisfy 
$\partial_\mu A^\mu\equiv 0$, and the classical 
equations of motion in vacuum are
given by $\partial^\nu\partial_\nu A^\mu=0$.  These follow,
for example, from the Hamiltonian \cite{schweber} 
\begin{equation} \label{emham}
H_{GB}[A,\pi] =
(1/2)\int dx\, \eta_{\mu\nu}\left(\pi^\mu\pi^\nu - \nabla A^\mu\cdot\nabla
A^\nu\right) ,
\end{equation}
where $\eta_{\mu\nu}$ denotes the Minkowski tensor, $\pi^\mu$ denotes
the conjugate momentum density, and $\nabla$ denotes the spatial
derivative. Here $H_{GB}$ corresponds to the gauge-breaking 
Lagrangian $L=-(1/2)\int dx\,A^{\mu,\nu}A_{\mu,\nu}$, and is seen 
to have the quadratic form of Eq.~(\ref{hquad}) 
(with $K^{\mu\nu}_x\equiv\eta_{\mu\nu}/2$).  

The exact uncertainty approach therefore immediately implies, via the
Corollary of the previous section, that the evolution of a 
{\it quantum} ensemble of electromagnetic fields is described by
the Schr\"{o}dinger equation
\begin{equation}
i\hbar(\partial\Psi/\partial t) = H_{GB}[A,-i\hbar(\delta/\delta A)]
\Psi ,
\end{equation} 
in agreement with the Gupta-Bleuler formalism \cite{schweber}.

Further, note that the probability of a member of the classical 
ensemble not satisfying the
Lorentz gauge condition $\partial_\mu A^\mu\equiv 0$ is zero by
assumption, i.e., the Lorentz gauge is equivalent to the condition
that the product $(\partial_\mu A^\mu) P[A]$ vanishes for all
physical fields.  For the quantum ensemble to satisfy this 
condition, as per the guiding principle discussed at the end of
Sec.~3  above, one equivalently requires,
noting Eq. (\ref{hdef}), that 
\[
(\partial_\mu A^\mu) \Psi[A] =0 .\]
As is well known, this constraint, if initially satisfied, is satisfied
for all times \cite{dirac} (as is the weaker constraint that
only the 4-divergence of the positive frequency part of the field vanishes
\cite{schweber}). Hence the evolution of the quantum
ensemble is consistent with the Lorentz gauge.
It would be of interest to derive the consistency of this constraint
directly from the equations of motion, Eqs. (\ref{modcont}) and
(\ref{modhj}) of Appendix C, for $P$ and $S$.

\subsection{Radiation gauge}
 
It is well known that one  can 
also obtain the classical equations of motion for the electromagnetic
field via an
alternative Hamiltonian, obtained 
by exploiting the degree of freedom left by the
Lorentz gauge to remove a dynamical 
coordinate (corresponding to the longitudinal polarisation).  In
particular, since $\partial_\mu A^\mu$ is invariant under 
$A^\mu\rightarrow A^\mu +\partial^\mu \chi$ for any function $\chi$
satisfying $\partial^\nu\partial_\nu\chi=0$, one may completely
fix the gauge in a given Lorentz frame by choosing $\chi$ such that
$A^0=0$.  One thus obtains, writing $A^\mu\equiv(A^0,{\bf A})$, the
radiation gauge $A^0=0$, $\nabla\cdot{\bf A}=0$.  

The classical
equations of motion, $\partial^\nu\partial_\nu{\bf A}=0$, follow 
from the Hamiltonian  
\begin{equation} \label{hc}
H_R[{\bf A},{\bm \pi}] = (1/2)\int dx\, ({\bm\pi }\cdot{\bm\pi }/\epsilon_0 +
|\nabla\times{\bf A}|^2/\mu_0 ) ,
\end{equation}
where ${\bm \pi}$ denotes the conjugate momentum density. Here $H_R$
corresponds to the standard Lagrangian $L=-(1/4\mu_0)
\int dx\,F^{\mu\nu}F_{\mu\nu}$,
with  $A^0\equiv 0$.  

This Hamiltonian has the quadratic
form of Eq. (\ref{hquad}), and hence the exact uncertainty approach 
yields the corresponding Schr\"{o}dinger equation
\begin{equation}
i\hbar(\partial\Psi/\partial t) = H_R[{\bf A},-i\hbar(\delta/\delta {\bf
A})]\Psi
\end{equation} 
for a quantum ensemble of electromagnetic 
fields in the radiation gauge  (this is also
the form of the Schr\"{o}dinger equation obtained via the
Schwinger-Tomonaga formalism \cite{wheeler}).  

Note that the electric field follows via Eqs.~(\ref{hc}) and
(\ref{conj}) as 
\[
{\bf E} = -\partial {\bf A}/\partial t = -\delta H_R/\delta{\bm \pi} =
-{\bm \pi}/\epsilon_0 , \]
and is therefore directly proportional to the classical momentum density ${\bm
\pi}$.  Fluctuations of the momentum density thus correspond to
fluctuations of the electric field {\bf E}.  
Further, the constraint $\nabla\cdot
{\bf A}=0$ implies there is a one-one relation between ${\bf A}$ and the
magnetic field ${\bf B}=\nabla\times{\bf A}$ (up to an additive
constant).  Uncertainty in the vector potential thus corresponds to
uncertainty in the magnetic field {\bf B}.
Hence, in the radiation gauge,
the exact uncertainty approach corresponds to adding nonclassical
fluctuations to the electric field components of an ensemble of 
electromagnetic fields, with the fluctuation
strength 
determined by the uncertainty in the magnetic field components.

\section{Example: Gravitational field}

\subsection{Hamilton-Jacobi constraints}

The gravitational field is described, up to arbitrary coordinate
transformations, by the metric tensor $g\equiv(g_{\mu\nu})$.  The corresponding
invariant length may be decomposed as \cite{dw}
\[
ds^2 = g_{\mu\nu}dx^\mu dx^\nu  = - (\alpha^2-{\bm \beta}\cdot {\bm \beta})dt^2
+ 2\beta_idx^idt+\gamma_{ij}dx^idx^j , \]
in terms of the lapse function $\alpha$, the shift function ${\bm
\beta}$, and the spatial 3-metric $\gamma\equiv(\gamma_{ij})$.  The
equations of motion are the Einstein field equations, which follow from
the Hamiltonian functional \cite{dw}
\begin{equation} \label{hdw}
H[\gamma, \pi,\alpha,{\bm \beta}] = \int dx\, \alpha{\cal H}_G[\gamma, \pi] -
2\int dx\,\beta_i\pi^{ij}_{~~|j}~ ,
\end{equation}
where $\pi\equiv(\pi^{ij})$ denotes the momentum density conjugate to
$\gamma$, $|j$ denotes the covariant 3-derivative, and the Hamiltonian
density ${\cal H}_G$ is given by
\begin{equation} \label{hg}
{\cal H}_G = 
(1/2) G_{ijkl}[\gamma]\pi^{ij}\pi^{kl} - 2\,
{}^{(3)}\!R[\gamma](\det\gamma)^{1/2} .
\end{equation}
Here ${}^{(3)}\!R$ is the curvature scalar corresponding to
$\gamma_{ij}$, and
\[
G_{ijkl}[\gamma] =(\gamma_{ik}\gamma_{jl}+\gamma_{il}\gamma_{jk}
-\gamma_{ij}\gamma_{kl})(\det \gamma)^{-1/2} . \]

The Hamiltonian functional $H$ corresponds to the standard Lagrangian 
$L=\int dx\,(-\det
g)^{1/2} R[g]$, 
where the momenta $\pi^0$ and $\pi^i$ conjugate to $\alpha$ and $\beta_i$
respectively vanish identically.  However, the lack of dependence of
$H$ on $\pi^0$ and $\pi^i$ is consistently maintained only if the rates
of change of these momenta also vanish, i.e., noting Eq. (\ref{conj}) of
Appendix B, only if the constraints \cite{dw}
\begin{equation} \label{constraint}
\delta H/\delta\alpha = {\cal H}_G = 0,~~~~\delta H/\delta\beta_i = -2
\pi^{ij}_{~~| j}=0
\end{equation}
are satisfied.  Thus the dynamics of the field
is independent of $\alpha$ and ${\bm \beta}$, so that these 
functions may be fixed arbitrarily.  Moreover, these
constraints 
immediately yield $H=0$ in Eq. (\ref{hdw}), and hence 
the system is static, with no explicit time dependence.

It follows that, in the {\it Hamilton-Jacobi} formulation of the 
equations of motion (see Appendix B), 
the momentum potential $S$ is 
independent of $\alpha$, ${\bm \beta}$ and $t$.  Noting that
$\pi\equiv\delta S/\delta\gamma$ in this formulation, Eqs.
(\ref{constraint}) therefore yield the corresponding constraints
\begin{equation} \label{cons}
\frac{\delta S}{\delta \alpha} = \frac{\delta S}{\delta\beta_i} = 
\frac{\partial S}{\partial t} = 0,~~~~
\left(\frac{\delta S}{\delta\gamma_{ij}}\right)_{| j}=0  
\end{equation} 
for $S$.  As shown by Peres \cite{peres}, a given functional $F[\gamma]$ of the
3-metric is invariant under spatial coordinate transformations if and
only if $(\delta 
F/\delta\gamma_{ij})_{|j}=0$, and hence the fourth constraint
in Eq.~(\ref{cons}) is equivalent to the invariance of $S$ under such
transformations.  This fourth constraint moreover implies that the
second 
term in Eq.~({\ref{hdw}) may be dropped from the Hamiltonian, 
yielding the 
reduced Hamiltonian 
\begin{equation} \label{hgr}
H_G[\gamma,\pi,\alpha] = \int dx\,\alpha {\cal H}_G[\gamma,\pi] 
\end{equation}
in the Hamiltonian-Jacobi formulation \cite{peres, gerlach}.

For an {\it ensemble} of classical gravitational fields, the 
independence of the dynamics with respect to $\alpha$, ${\bm \beta}$ and 
$t$ implies that members of the ensemble are distinguishable only by their
corresponding 3-metric $\gamma$.  Moreover, it 
is natural to impose the additional geometric requirement that the ensemble is
invariant under spatial coordinate transformations.  One therefore has
the constraints
\begin{equation} \label{conp}
\frac{\delta P}{\delta\alpha}= \frac{\delta P}{\delta\beta_i} =
\frac{\partial P}{\partial t} = 0,~~~~
\left(\frac{\delta P}{\delta\gamma_{ij}}\right)_{| j}=0 
\end{equation}
for the corresponding probability density functional $P[\gamma]$,
analogous to Eq. (\ref{cons}). The first two constraints imply that
ensemble averages only involve integration over $\gamma$.

\subsection{Quantum ensembles and operator-ordering}
 
Noting Eq.~(\ref{hg}), the Hamiltonian $H_G$ in Eq.~(\ref{hgr}) 
has the quadratic form of Eq. (\ref{hquad}). Hence 
the exact uncertainty approach is applicable, and immediately leads to the
Schr\"{o}dinger equation 
\begin{equation} \label{seq}
i\hbar\partial\Psi/\partial t=\int dx\,\alpha {\cal H}_G[\gamma,
-i\hbar(\delta/\delta\gamma)]\Psi 
\end{equation}
for a {\it quantum} ensemble of gravitational fields, as per the
Corollary of Sec.~3.  

As discussed at the end of Sec.~3, we follow the guiding
principle that all constraints imposed on the classical ensemble should
be carried over to corresponding constraints on the quantum ensemble.  
Thus, from Eqs.~(\ref{cons}) and (\ref{conp}) we require that $P$ and
$S$, and hence $\Psi$ in Eq.~(\ref{hdef}), are independent of $\alpha$,
${\bm \beta}$ and $t$ and invariant under spatial coordinate
transformations, i.e., 
\begin{equation} \label{conpsi}
\frac{\delta \Psi}{\delta\alpha}= \frac{\delta \Psi}{\delta\beta_i} =
\frac{\partial \Psi}{\partial t} = 0,~~~~
\left(\frac{\delta \Psi}{\delta\gamma_{ij}}\right)_{| j}=0 .
\end{equation}
Applying the first and third of these constraints to Eq.~(\ref{seq})
immediately yields, via Eq.~(\ref{hg}),  
the reduced Schr\"{o}dinger equation
\begin{equation} \label{wdw}
{\cal H}_G[\gamma, -i\hbar(\delta/\delta \gamma)]\Psi = (-\hbar^2/2)
\frac{\delta}{\delta \gamma_{ij}}G_{ijkl}[\gamma]\frac{\delta
}{\delta \gamma_{kl}}\Psi - 2\,{}^{(3)}\!R[\gamma](\det\gamma)^{1/2}
\Psi = 0 ,
\end{equation} 
which may be recognised as  
the Wheeler-DeWitt equation in the metric representation \cite{dw}. 

A notable feature of Eq.~(\ref{wdw}) is that the Wheeler-DeWitt equation
has not only been derived from an exact uncertainty principle: it has, as 
a consequence of Eq.~(\ref{corollary}),
been derived with a {\it precisely}
defined operator ordering (with $G_{ijkl}$ sandwiched between the two
functional derivatives).  
Thus the exact uncertainty approach does not admit
ambiguity in this respect, unlike the standard
approach \cite{dw}.  Such removal of ambiguity is essential to
making definite physical predictions, and hence may be
regarded as an advantage of the exact uncertainty approach. 

For example, Kontoleon and Wiltshire \cite{wiltshire} have pointed
out that Vilenkin's prediction of inflation in minisuperspace,
from a corresponding Wheeler-DeWitt equation with
``tunneling'' boundary conditions \cite{vilenkin},
depends critically upon the operator ordering used.
In particular, considering the class of orderings defined by an integer
power $p$, with corresponding
Wheeler-DeWitt equation \cite{vilenkin}
\begin{equation} \label{vile}
\left[\frac{\partial^2}{\partial a^2} +
\frac{p}{a}\frac{\partial}{\partial a} - \frac{1}{a^2}\frac{\partial^2}{
\partial \phi^2} - U(a,\phi)\right]\Psi = 0 
\end{equation}
(for a Friedmann-Robertson-Walker metric coupled to a scalar field $\phi$), 
Kontoleon and Wiltshire show that Vilenkin's approach fails for orderings
with $p\geq 1$ \cite{wiltshire}.  Moreover, they suggest
that the only natural ordering is in fact the ``Laplacian'' ordering
corresponding to $p=1$, which has been justified on geometric grounds by
Hawking and Page \cite{hawking}.  

However, noting that the relevant Hamiltonian functional in Eq.~(2.7) of
Ref.~\cite{vilenkin} is quadratic in the momentum densities of the
metric and the scalar field, the exact uncertainty approach may be
applied, and yields the
Wheeler-DeWitt equation corresponding to $p=-1$ in Eq. (\ref{vile}). 
Hence the criticism in Ref. \cite{wiltshire} is avoided.  
One also has the nice feature
that  the associated Wheeler-DeWitt equation
can be exactly solved for this ``exact uncertainty'' ordering \cite{vilenkin}.

A certain degree of ambiguity remains, which derives from the need to
introduce some sort of regularisation scheme to remove divergences
arising from the product of two functional derivatives acting at the
same point in the Wheeler-DeWitt equation.  Such considerations,
however, do not play a role in the example that we have just discussed,
which concerns minisuperspace quantisation involving a finite number of
degrees of freedom.  It is important to distinguish this regularisation
problem from the far more difficult one associated with the requirement
of ``Dirac consistency'', i.e., the need to find a choice of operator
ordering {\it and} regularisation scheme that will permit mapping the
classical Poisson bracket algebra of constraints to an algebra of
operators within the context of the Dirac quantisation of canonical
gravity \cite{tsamis}.  Our approach is based on the
Hamilton-Jacobi formulation of classical gravity and, as shown by
Bergmann \cite{bergmann}, the functional form of the Hamilton-Jacobi
functional $S$ is already invariant under the action of the group
generated by the constraints.

Finally, note that a similar approach may be applied to the Ashtekar
formalism for gravitational fields \cite{ashtekar}, where again the
Hamiltonian is quadratic in the field momentum density (in particular,
the two constraints linear in the momentum density $\tilde{\sigma}$
become constraints on the Hamilton-Jacobi functional $S$, corresponding
to the invariance of $S$ under spatial coordinate and internal gauge
transformations \cite{rovelli}, while the constraint quadratic in
$\tilde{\sigma}$ generates the Ashtekar-Wheeler-DeWitt equation with a
unique operator ordering).

\section{Discussion}

The main result of this paper is the derivation of the 
quantum equation of motion, Eq. (\ref{corollary}), from an exact
uncertainty principle, for fields with
Hamiltonian functionals quadratic in the momentum density.

It is important to emphasise that the exact uncertainty approach 
does {\it not} assume the existence of a complex amplitude functional
$\Psi[f]$, nor the representation of fields by operators, nor the
existence of a universal constant $\hbar$ with units of action, nor the
existence of a linear operator equation in some Hilbert space.  
Only the assumptions of
causality, independence, invariance and exact uncertainty are required, all
formulated in terms of a {\it single}
nonclassical element (the uncertainty 
introduced by the momentum fluctuation $N$).  Since uncertainty is at
the conceptual core of quantum mechanics, this is an elegant and
pleasing result.

The assumptions used also provide an intuitive picture for the origin of the
Schr\"{o}dinger equation for bosonic fields, 
as arising from nonclassical fluctuations of
the momentum density.  Of course this picture has limitations - the
fluctuations arise from the uncertainty of the field itself, rather
than from some external source, and hence are 
most certainly ``nonclassical'' rather than ``classical'' 
in nature.  

A minimalist interpretation of the exact uncertainty approach, based on
Eqs. (\ref{noise}) and (\ref{theorem}), is as follows.
Every {\it physical} field has an intrinsic uncertainty, 
which is modelled by a corresponding statistical ensemble.  
Further, the nature of
this inherent uncertainty is such as to preclude
a deterministic relationship between the field and its
conjugate momentum density - one must introduce fluctuations into the
classical relationship, as per Eq.~(\ref{noise}).  However, the {\it degree} of indeterminism in
this relationship {\it is} precisely quantifiable, in a statistical sense,
being directly specified by the 
inherent field uncertainty as per Eq.~(\ref{theorem}).  

The above interpretation may be regarded as a significant sharpening of
the so-called ``statistical interpretation'' of quantum mechanics
\cite{ballentine}, and is notably very different to
the ``causal interpretation'' of Bohm and co-workers \cite{holland}.
In the latter it is assumed that there is a pre-existing complex
amplitude functional $\Psi[f]=\sqrt{P}\exp(iS/\hbar)$ 
obeying a Schr\"{o}dinger equation, 
which acts upon a single classical field  via
the addition of a ``quantum potential", $Q[P]$, to the classical Hamiltonian.
It is further assumed that the momentum density is precisely 
$g\equiv\delta S/\delta f$, and that physical ensembles of fields have 
probability density functional $P=|\Psi|^2$.
In contrast, the exact
uncertainty approach does not postulate the existence of adjunct
amplitudes and potentials; the Schr\"{o}dinger equation 
directly represents the evolution of an ensemble, rather than of an 
external amplitude functional acting on individual systems 
(and is derived rather than postulated); 
and the basic tenet in Eq. (\ref{noise}) 
is that $g\neq\delta S/\delta f$.

A further strength of the exact uncertainty approach is
that the basic underlying concept - the addition of ``nonclassical''
momentum
fluctuations to a classical ensemble - carries through from quantum 
particles to quantum fields without creating 
conceptual difficulties. This adds an interpretational 
strength to the exact uncertainty
approach not mirrored in other approaches
that rely on connecting the equations of motion of
classical and quantum ensembles.  For example, the abovementioned ``causal
interpretation'' of Bohm and co-workers
is explicity non-local, and hence
{\it non}-causal, for relativistic fields \cite{holland}. As another
example, one cannot
simultaneously describe both the electric and the magnetic fields in
generalisations of Nelson's stochastic approach to electromagnetic
fields \cite{nelson}.

It is of interest to consider the scope and limitations of the exact
uncertainty approach to physical systems.  
This approach has previously been applied to quantum particles
\cite{hallreg, bamberg}, and may be generalised to obtain the
Pauli equation for a non-relativistic spin-1/2 particle, 
and the Schr\"{o}dinger equation for
particles with
position-dependent mass  (where in the latter case one obtains the 
unique ordering $\hat{p}[2m(\hat{x})]^{-1}\hat{p} +
V(\hat{x})$ for the Hamiltonian operator, 
corresponding to the ordering
parameter $\alpha=0$ in Ref.~\cite{plastino}). 
In this paper the approach has been further generalised
to bosonic quantum fields with Hamiltonians quadratic in the momentum
density (including all relativistic integer-spin fields). 
It is also, indirectly, applicable
to the {\it non}-quadratic Hamiltonian functional
of a nonrelativistic Schr\"{o}dinger field (corresponding to second quantisation
of the
particle Schr\"{o}dinger equation), in the sense that this
field may be obtained as a low-energy limit of the complex Klein-Gordon
field
\cite{brown} (to which the exact uncertainty approach directly applies).

However, a major question to be addressed in the future is whether the exact
uncertainty approach is applicable to the derivation of fermionic field
equations. These have
two features which present challenges: the corresponding
ensemble Hamiltonian is usually linear in the momentum density, and the
anticommutation relations make it difficult to connect the equations of
motion with corresponding classical equations of motion in the limit as
$\hbar\rightarrow 0$.  One possible approach is to determine whether
exact uncertainty relations exist for such fields, analogous to Eq.
(\ref{eur}), 
as these might suggest the statistical
properties required by suitable ``nonclassical'' fluctuations
(note that the exact
uncertainty relations satisfied by bosonic fields are derived
in Ref.~\cite{eprint}).

Finally, in this paper the basic Schr\"{o}dinger equation for bosonic fields
has been obtained, with the advantageous features of an intuitive
picture for the origin of the ``quantum'' nature of such fields, and a 
unique operator ordering in cases where other approaches are ambiguous.
It would be of interest to consider further issues, such as the
representation of general physical observables by operators
(addressed for the case of particles in Ref. \cite{hallreg}), boundary
conditions, infinities, etc, from the new perspective on the conceptual
and technical basis of
quantisation offered by the exact uncertainty approach.

\appendix
\section{Functional derivatives and integrals}
The necessary definitions and properties of functionals are noted here,
including variational properties of functional integrals.

A functional, $F[f]$, is a mapping from a set of physical fields
(assumed to form a vector space) to the real or complex numbers, and the
functional derivative of $F[f]$ is defined via the variation of $F$ with
respect to $f$, i.e., 
\begin{equation} \label{fderiv}
\delta F= 
F[f+\delta f] - F[f] = \int dx\, \frac{\delta F}{\delta f_x}\, \delta\!f_x
\end{equation}
for arbitrary infinitesimal variations $\delta f$.  Thus the functional
derivative is a 
field density, $\delta F/\delta f$, having the value $\delta
F/\delta f_x$ at position $x$.  For curved spaces one may explicitly 
include a volume element in the integral, thus redefining the functional
derivative by a multiplicative function of $x$; however, this is 
merely a matter of taste
and will not be adopted here.  The functional derivative is
assumed to always exist for the functionals in this paper.

It follows directly from Eq.~(\ref{fderiv}) that the functional
derivative satisfies product and chain rules analogous to ordinary
differentiation. The choice $F[f] = f_{x'}$ in Eq.
(\ref{fderiv}) yields
$\delta f_{x'}/\delta f_x = \delta(x-x')$.
Moreover, if the field depends on some parameter, $t$ say, then
choosing $\delta f_x = f_x(t+\delta t) - f_x(t)$ in Eq. (\ref{fderiv})
yields
\begin{equation} \label{rate}
\frac{dF}{dt} = \frac{\partial F}{\partial t} + \int dx\,\frac{\delta
F}{\delta f_x} \frac{\partial f_x}{\partial t} 
\end{equation}
for the rate of change of $F$ with respect to $t$.

Functional integrals correspond to integration  of functionals over the
vector space of physical fields (or equivalence classes thereof).  
The only property we require for this paper is the existence of a measure
$D\!f$ on this vector space which is {\it translation invariant}, i.e.,
$\int D\!f \equiv \int D\!f'$ for any translation $f' = f + h$ (which 
follows immediately, for example, from the discretisation approach to
functional integration \cite{brown}). In particular, this 
property implies the useful result 
\begin{equation} \label{div}
\int D\!f\,\frac{\delta F}{\delta f} = 0~~~{\rm for}~~~\int D\!f\,
F[f] < \infty ,
\end{equation}
which is used repeatedly below and in the text.  Eq. (\ref{div}) follows
by noting that the finiteness condition and translation invariance imply
\[ 0 = \int D\!f\, (F[f+\delta f] - F[f]) = \int dx\,\delta\!f_x \left(
\int D\!f\,
\delta F/\delta f_x\right) \]
for arbitrary infinitesimal translations.

Thus, for example, if $F[f]$ has a finite expectation value
with respect to some probability density functional $P[f]$, then 
Eq. (\ref{div}) yields the ``integration by parts'' formula
\[
\int D\!f\,P (\delta F/\delta f) = -\int D\!f\, (\delta P/\delta f)F .\]
Moreover, from Eq. (\ref{div}) the total probability,
$\int D\!f\, P$,
is conserved for any probability flow satisfying a continuity equation of
the form
\begin{equation} \label{gcont}
\frac{\partial P}{\partial t} + \int dx\, \frac{\delta}{\delta f_x} \left[ 
PV_x\right] = 0 ,
\end{equation}
providing that the average flow rate, $\langle V_x\rangle$, is finite. 

Finally, consider a functional integral of the form
\begin{equation} \label{mdensity}
I[F] = \int D\!f\,\xi(F, \delta F/\delta f) ,
\end{equation}
where $\xi$ denotes any function of some functional $F$ and its functional
derivative.  Variation of $I[F]$ with respect to $F$ then gives, to
first order,

\begin{eqnarray*}
\Delta I = I[F+\Delta F]-I[F] & = & \int D\!f \,\left\{
(\partial \xi/\partial F)\Delta F +\int dx\, \left[
\partial \xi/\partial (\delta F/\delta f_x)\right]\left[\delta
(\Delta F)/\delta f_x\right]\right\}\\
& = & \int D\!f\, \left\{(\partial \xi/\partial F) - \int dx\,
\frac{\delta}{\delta f_x} \left[ \partial \xi/\partial (\delta
F/\delta f_x)\right]\right\}\Delta F\\
& & \mbox{} + \int dx\,\int D\!f\, \frac{\delta}{\delta f_x} \left\{
\left[ \partial \xi/\partial (\delta F/\delta f_x)\right]\Delta F
\right\} .
\end{eqnarray*}
Assuming that the functional integral of the expression in curly
brackets in the last term is finite, this term vanishes from Eq.
(\ref{div}), yielding  the result
\[ \Delta I = \int D\!f\, \frac{\Delta I}{\Delta F}\, \Delta F \]
analogous to Eq.~(\ref{fderiv}), where the variational derivative
$\Delta I/\Delta F$ is defined by
\begin{equation} \label{mform}
\frac{\Delta I}{\Delta F} := \frac{\partial \xi}{\partial F}
- \int dx\,\frac{\delta}{\delta f_x}\left[\frac{\partial \xi}{
\partial (\delta F/\delta f_x)}\right] .
\end{equation}
A similar result holds for multicomponent fields, with summation over
the discrete index $a$ in the second term.

\section{ Hamilton-Jacobi ensembles}
The salient aspects of the Hamilton-Jacobi formulation of classical
field theory \cite{goldstein}
are collected here, with particular attention to the origin
of the associated continuity equation for {\it ensembles} of classical
fields, required in Secs.~2 and 3.

Two classical fields $f$, $g$ are canonically conjugate if there is a
Hamiltonian functional $H[f,g,t]$ such that 
\begin{equation} \label{conj}
\partial f/\partial t = \delta H/\delta g, ~~~~\partial g/\partial t =
-\delta H/\delta f .
\end{equation}
These equations follow from the action principle $\delta A = 0$, with
action functional $A = \int dt\,[-H + \int dx\,g_x(\partial f_x/\partial
t)]$.  The rate of change of an arbitrary functional
$G[f,g,t]$ follows from Eqs. (\ref{rate}) and (\ref{conj}) as
\[
\frac{dG}{dt} = \frac{\partial G}{\partial t} + \int dx\,\left(
\frac{\delta G}{\delta f_x} \frac{\delta H}{\delta g_x} - \frac{\delta
G}{\delta g_x}\frac{\delta H}{\delta f_x} \right)
=: \frac{\partial G}{\partial
t} + \{ G,H\} , 
\]
where $\{~,~\}$ is a generalised Poisson bracket.

A canonical transformation maps $f$, $g$ and $H$ to $f'$, $g'$ and $H'$,
such that the equations of motion for the latter retain the canonical
form of Eq. (\ref{conj}).  Equating the variations of the corresponding
actions $A$ and $A'$ to zero, it follows that all physical trajectories
must satisfy
\[
-H + \int dx\, g_x(\partial f_x/\partial t) = -H' + \int dx\,g_x'(\partial
f_x'/\partial t) + dF/dt \]
for some ``generating functional" $F$.
Now, any two of the fields $f,g, f', g'$ determine the remaining two fields for
a given canonical transformation.  Choosing $f$ and $g'$ as the two
independent fields, defining the new generating functional $G[f,g',t] =
F + \int dx\, f_x'g_x'$, and using Eq.~(\ref{rate}), then yields 
\[
H' = H + \frac{\partial G}{\partial t} + \int dx\, \left[ \frac{\partial
f_x}{\partial t}\left( \frac{\delta G}{\delta f_x} -g_x\right) +
\frac{\partial g_x'}{\partial t}\left(\frac{\delta G}{\delta g_x'} -
f_x'\right)\right] \]
for all physical trajectories.  The terms in round brackets 
therefore vanish identically, yielding the generating relations
\begin{equation} \label{sgen}
H' = H + \partial G/\partial t,~~~~ g=\delta G/\delta
f,~~~~f'=\delta G/\delta g' .
\end{equation}
A canonical transformation is thus completely specified by the associated
generating functional $G$.

To obtain the {\it Hamilton-Jacobi} formulation of the  equations of
motion, consider a
canonical transformation to fields $f'$, $g'$ which are time-independent
(eg, to the fields $f$ and $g$ at some fixed time $t_0$). From 
Eq.~(\ref{conj}) one may  
choose the corresponding Hamiltonian $H'\equiv 0$ 
without loss of generality, and hence from 
Eq.~(\ref{sgen}) the momentum density and the associated
generating functional $S$ are specified by the functional equations
\begin{equation} \label{hj}
g=\frac{\delta S}{\delta f},~~~~
\frac{\partial S}{\partial t} + H[f, \delta S/\delta f,t] = 0 .
\end{equation}
The latter is the desired Hamilton-Jacobi equation. Solving this equation for
$S$ is equivalent to solving Eqs.~(\ref{conj}) for $f$ and $g$. 

Note that along a physical trajectory one has $g'\equiv$ constant, 
and hence from Eqs. (\ref{rate}) and
(\ref{hj}) that  
\[
\frac{dS}{dt} = \frac{\partial S}{\partial t} + \int dx\, \frac{\delta
S}{\delta f_x} \frac{\partial f_x}{\partial t} = -H + \int
dx\,g_x\frac{\partial f_x}{\partial t} = \frac{dA}{dt} . \]
Thus the Hamilton-Jacobi functional $S$ is equal to the action
functional $A$, up to an additive 
constant.  This relation underlies the connection
 between the derivation of the Hamilton-Jacobi equation from a
particular type of canonical transformation, as above, and the
derivation from a particular type of variation of the action, as per the
Schwinger-Tomonaga formalism \cite{wheeler,schwinger}.  

The Hamilton-Jacobi formulation has the interesting feature that once $S$
is specified, the momentum density is determined by the relation
$g=\delta S/\delta f$, i.e., it is a functional of $f$.  Thus, unlike
the Hamiltonian formulation of Eqs. (\ref{conj}), an {\it ensemble} of
fields is specified by a probability density functional $P[f]$, not by a
phase space density functional $\rho[f,g]$. 

In either case, the equation of motion for the probability density 
corresponds to the
conservation of probability, i.e., to a continuity equation as per Eq.
(\ref{gcont}). For example, in the Hamiltonian formulation 
the associated
continuity equation for $\rho[f,g]$ is 
\[ \partial \rho/\partial t + \int dx\,\{
(\delta/\delta f_x)[\rho(\partial f_x/\partial t)] + (\delta/\delta
g_x[\rho(\partial g_x/\partial t)]\} = 0 ,\]
which reduces to the Liouville
equation $\partial \rho/\partial t = \{H, \rho\}$ via Eqs. (\ref{conj}).

Similarly, in the Hamilton-Jacobi formulation, 
the rate of change of the field $f$ follows from Eqs. (\ref{conj})
and (\ref{hj}) as the functional 
\[
V_x[f] = \partial f_x/\partial t = (\delta H/\delta g_x)
\left|{}_{g=\delta S/\delta f}\right. ,\]
and hence the associated continuity equation for an ensemble of fields
described by $P[f]$ follows via Eq. (\ref{gcont}) as 
\begin{equation} \label{hjcont}
\frac{\partial P}{\partial t} + \int dx\, \frac{\delta}{\delta f_x}
\left[ P \left. \frac{\delta H}{\delta g_x}\right|_{g=\delta S/\delta f}
\right] . 
\end{equation}

Eqs. (\ref{hj}) and (\ref{hjcont}) generalise immediately to
multicomponent fields, and form the basis of the classical
starting point in the derivation of the quantum equations of motion for
bosonic fields in Secs.~2 and 3.

\section{Proofs of the Theorem and Corollary}

{\bf Proof of Theorem (Eq. \ref{theorem}):} 
From the causality and exact uncertainty assumptions in Section 3,
one has ${\rm Cov}_x(N) = \alpha(P, \delta P/\delta f_x, f_x)$.  To
avoid issues of regularisation, it is convenient to consider a
position-dependent canonical transformation, $f_x\rightarrow
\Lambda_x^{-1}f_x$, such that $A[\Lambda]:=\exp[\int dx\,\ln
|\det\Lambda_x|]$ is
finite. Then the probability density functional
$P$ and the measure $D\!f$ transform as $P\rightarrow AP$ and
$D\!f\rightarrow A^{-1}D\!f$ respectively, and so the
invariance assumption in Section 3 requires that
\[
\alpha(AP, A\Lambda_x^Tu, \Lambda_x^{-1}w) \equiv
\Lambda_x^T\alpha(P,u,w)\Lambda_x , \]
where $u^a$ and $w^a$ denote the vectors $\delta P/\delta f^a_x$ and
$f^a_x$ respectively, for a given value of $x$.
Since $\Lambda_x$ can remain the same at a given point $x$ while varying
elsewhere, this homogeneity condition must hold for $A$ and $\Lambda_x$
independently.  Thus, choosing $\Lambda_x$ to be the identity matrix at
some point $x$, one has $\alpha(AP, Au,w) = \alpha(P,u,w)$ for all $A$,
implying that $\alpha$ can involve $P$ only
via the combination $v:=u/P$.

The above homogeneity condition for $\alpha$ therefore reduces to
\[
\alpha(\Lambda^Tv, \Lambda^{-1}w) = \Lambda^T\alpha(v,w)\Lambda~.\]
Note that this equation is linear, and invariant under multiplication of
$\alpha$ by any function of the scalar $J:=v^Tw$.  Moreover, it may
easily be checked that if $\sigma$ and $\tau$ are solutions, then so are
$\sigma \tau^{-1}\sigma$ and $\tau\sigma^{-1}\tau$. Choosing the two
independent solutions $\sigma=vv^T$, $\tau=(ww^T)^{-1}$, it follows that
the general solution has the form
\[
\alpha(v,w) = \beta(J)vv^T + \gamma(J)(ww^T)^{-1} \]
for arbitrary functions $\beta$ and $\gamma$.

For $P=P_1P_2$ one finds $v=(v_1,v_2)$, $w=(w_1,w_2)$, where the
subscripts label corresponding subensemble quantities, and hence the
independence assumption in Section 3 reduces to the requirements
\[
\beta(J_1+J_2)v_1v_1^T + \gamma(J_1+J_2)(w_1w_1^T)^{-1} =
\beta_1(J_1)v_1v_1^T + \gamma_1(J_1)(w_1w_1^T)^{-1}, \]
\[
\beta(J_1+J_2)v_2v_2^T + \gamma(J_1+J_2)(w_2w_2^T)^{-1} =
\beta_2(J_2)v_2v_2^T + \gamma_2(J_2)(w_2w_2^T)^{-1} ,\]
for the respective subensemble covariance matrices.
Thus $\beta=\beta_1=\beta_2=C$, $\gamma=\gamma_1=\gamma_2=D$ for
universal (i.e., system-independent)
constants $C$ and $D$, yielding the general form
\[
\left[{\rm Cov}_x(N)\right]^{ab} = C (\delta P/\delta f^a_x) (\delta
P/\delta f^b_x)/P^2 + D W^{ab}_x[f] \]
for the fluctuation covariance matrix, where $W_x[f]$ denotes the
inverse of
the matrix with ${ab}$-coefficient $f^a_xf^b_x$. 

Since $W_x[f]$ is
purely a functional of $f$, it merely contributes a classical
additive potential term to the ensemble Hamiltonian of Eq.
(\ref{hprime}).  It
thus has no nonclassical role, and can be absorbed directly into the
classical potential $\langle V\rangle$ (indeed, for fields with more
than one
component this term is singularly ill-defined, and hence can be
discarded on physical grounds).  Thus we may take $D=0$ without loss of
generality. Finally, the positivity of $C$ follows from the positivity
of the covariance matrix ${\rm Cov}_x(N)$, and the theorem is proved.
 
{\bf Proof of Corollary (Eq. \ref{corollary}):}
First, the equations of motion corresponding
to the ensemble Hamiltonian $\tilde{H}'$ follow via the theorem and
Eqs.~(\ref{mconj}) as: (a) the continuity
equation Eq. (\ref{mcont}) as before (since the additional term does not
depend on $S$), which from Eq. (\ref{hquad}) has the explicit form
\begin{equation} \label{modcont}
\frac{\partial P}{\partial t} + 2\sum_{a,b}\int
dx\,\frac{\delta}{\delta f^a_x}\left( PK^{ab}_x\frac{\delta S}{\delta
f^b_x}\right) = 0 ;
\end{equation}
and (b) the modified Hamilton-Jacobi equation
\[ \partial S/\partial t = -\Delta\tilde{H}'/\Delta P = -H[f,\delta
S/\delta f,t] - \Delta(\tilde{H}'-\tilde{H})/\Delta P .\]
Calculating the last term via Eq. (\ref{theorem}) and
Eq. (\ref{mform}) of Appendix A, this
simplifies to
\begin{equation} \label{modhj}
\frac{\partial S}{\partial t}+H[f, \delta S/\delta f,t]
-4CP^{-1/2} \sum_{a,b}
\int dx\,\left(K^{ab}_x\frac{\delta^2 P^{1/2}}{\delta f^a_x \delta
f^b_x}
+\frac{\delta K^{ab}_x}{\delta f^a_x}\frac{\delta P^{1/2}}{\delta f^b_x}
\right) =0 .
\end{equation}

Second, writing $\Psi=P^{1/2}\exp(iS/\hbar)$, multiplying each side of
Eq. (\ref{corollary})
on the left by $\Psi^{-1}$, and expanding, gives a complex equation for
$P$ and $S$. The imaginary part is just the continuity
equation of Eq. (\ref{modcont}), and the real part is the modified
Hamilton-Jacobi equation of Eq. (\ref{modhj}) above, providing that one
identifies $C$ with $\hbar^2/4$.


\begin{thebibliography}{99}
\bibitem{eur} Hall MJW 2001 {\it Phys. Rev. A} {\bf 64} 052103 
\bibitem{hallreg} Hall MJW and Reginatto M 2002 {\it J. Phys. A} {\bf 35} 3289
\bibitem{bamberg} Hall MJW and Reginatto M 2002 {\it Fortschr. Phys.} {\bf
50} 646; reprinted in Papenfuss D, L\"{u}st D and Schleich WP (editors) 2002
{\it 100 Years Werner Heisenberg: Works and
Impact} (Wiley: Berlin)
\bibitem{vilenkin} Vilenkin A 1986 {\it Phys. Rev. D} {\bf 33} 3560
\bibitem{brown} Brown LS 1992 {\it Quantum Field Theory}
(Cambridge: Cambridge University Press), Chapters 1, 3
\bibitem{schweber} Schweber SS 1961 {\it An Introduction to Relativistic
Quantum Field Theory} (New York: Row, Peterson and Co.), Chapter 9
\bibitem{dirac} Dirac PAM 1966 {\it Lectures on Quantum Field Theory}
(New York: Academic), Chapters 14, 15 
\bibitem{wheeler} Wheeler JA 1970 in {\it Analytical Methods in
Mathematical Physics} edited by Gilbert RP and Newton RG (New York: Gordon
and Breach), p.~335
\bibitem{dw} De Witt BS 1967 {\it Phys. Rev.} {\bf 160} 1113 
\bibitem{peres} Peres A 1962 {\it Nuovo Cim.} {\bf 26} 53 
\bibitem{gerlach} Gerlach UH 1969 {\it Phys. Rev.} {\bf 177} 1929 
\bibitem{wiltshire} Kontoleon N and Wiltshire DL 1999 {\it Phys. Rev. D} {\bf
59} 063513 
\bibitem{hawking} Hawking SW and Page DN 1986 {\it Nucl. Phys.} {\bf B264} 185
\bibitem{tsamis} Tsamis NC and Woodard RP 1987 {\it Phys. Rev D} {\bf 36}
3641
\bibitem{bergmann} Bergmann PG 1966 {\it Phys. Rev.} {\bf 144} 1078
\bibitem{ashtekar} Ashtekar A 1986 {\it  Phys. Rev. Lett.} {\bf 57} 2244 
\bibitem{rovelli} Rovelli C 2002 eprint gr-qc/0207043
\bibitem{ballentine} Ballentine LE 1970 {\it Rev. Mod. Phys.} {\bf 42} 358 
\bibitem{holland} Holland PR 1993 {\it The Quantum Theory of Motion}
(Cambridge: Cambridge University Press), sections 12.4, 12.5
\bibitem{nelson} Davidson M 1981 {\it J. Math. Phys.} {\bf 22} 2588
\bibitem{plastino} Plastino AR, Casas M and Plastino A 2001 {\it Phys. Lett.
A} {\bf 281} 297 
\bibitem{eprint} Hall MJW, Kumar K and Reginatto M 2002 eprint hep-th/0206235
\bibitem{goldstein} Goldstein H 1950 {\it Classical Mechanics}
(New York: Addison-Wesley), Chapters 8, 9, 11
\bibitem{schwinger} Roman P 1969 {\it Introduction to Quantum Field
Theory} (New York: Wiley), Chapter 2
\end{thebibliography}
\end{document}